\documentclass[twocolumn,apl,amsmath,amssymb,showpacs,superscriptaddress]{revtex4-2}
\usepackage{epsf}
\usepackage[utf8]{inputenc}
\usepackage{amsmath}
\usepackage{amsfonts}
\usepackage{amssymb}
\usepackage{makeidx}
\usepackage{color}
\usepackage{gensymb}
\usepackage{graphicx}

\begin{document}
\title {\normalsize Unveiling the magnetic structure and phase transition of Cr$_2$CoAl using neutron diffraction}
\author{Guru Dutt Gupt}
\affiliation{Department of Physics, Indian Institute of Technology Delhi, Hauz Khas, New Delhi-110016, India}
\author{Yousef Kareri}
\email{Present Address: Saudi Electronic University, Saudi Arabia}
\affiliation{School of Physics, The University of New South Wales, Kensington, 2052 NSW, Sydney, Australia}
\author{James Hester}
\affiliation{Australian Center for Neutron Scattering, Australian Nuclear Science and Technology Organisation (ANSTO), New Illawarra Road, Lucas Heights NSW 2234, Australia}
\author{Clemens Ulrich}
\affiliation{School of Physics, The University of New South Wales, Kensington, 2052 NSW, Sydney, Australia}
\author{R. S. Dhaka}
\email{Corresponding author's email: rsdhaka@physics.iitd.ac.in}
\affiliation{Department of Physics, Indian Institute of Technology Delhi, Hauz Khas, New Delhi-110016, India}

\date{\today}
	
\begin{abstract}
We report the detailed analysis of temperature dependent neutron diffraction pattern of the Cr$_2$CoAl inverse Heusler alloy and unveil the magnetic structure up to the phase transition as well as its fully compensated ferrimagnetic nature. The Rietveld refinement of the diffraction pattern using the space group I$\bar4${\it m}2 confirm the inverse tetragonal structure over the large temperature range from 100~K to 900~K. The refinement of the magnetic phase considering the wave vector $k=$ (0, 0, 0) reveals the ferrimagnetic nature of the sample below 730$\pm$5~K. This transition temperature is obtained from empirical power law fitting of the variation in the ordered net magnetic moment and intensity of (110) peak as a function of temperature. The spin configuration of the microscopic magnetic structure suggests the nearly fully compensated ferrimagnetic behavior where the magnetic moments of Cr2 are antiparallel with respect to the Cr1, and Co moments. Moreover, the observed anomaly in the thermal expansion and lattice parameters at 730$\pm$5~K suggest that the distortion in crystal structure may play an important role in the magnetic phase transition.  
\end{abstract}
\maketitle

\section{\noindent ~Introduction}
	
	In the field of spin filters and spintronics, the compensated ferrimagnetic and spin gapless semiconductor Heusler alloys emerge as potential candidates because of their exotic physical properties elegantly controlled by the conduction method of the electrons at the Fermi level (E$_{\rm F}$) \cite{Felser_springer_16, Zutic_rmp_04}. More interestingly, a few Heusler alloys show half-metallic (HM) nature with 100\% spin polarization where the conduction is only due to one spin channel and there are no electrons present with opposite spin at E$_{\rm F}$ \cite{Groot_prl_83}. Leuken and de Groot \cite{Leuken_prl_95} theoretically proposed to realize HM antiferromagnets (AFMs) with full spin polarization in Heusler alloys, which are defined as zero net moments with the fully spin polarization, and are also classified as the HM fully compensated ferrimagnets (FCF) \cite{Gao_apl_13, Hakimi_jap_13}. These type of HM-AFM/FCF materials have advantage due to their zero stray magnetic field and therefore no energy losses during device operation for vaious applications \cite{Gao_apl_13}. In recent time, inverse Heusler (X$_2$YZ)  alloys, where the atomic number of X is smaller than Y and crystal structure changes from L2$_1$ to XA type, are predicted to show such vital magnetic properties as well as spin gapless semiconducting nature, and therefore are considered as potential candidates for practical applications \cite{Gasi_apl_13, SkaftourosAPL13, MohantaJMMM17, WinterlikAM12, ZhangEPL15, GalanakisPRB07, GalanakisJPCM14}. In this family, Cr$_2$CoAl was theoretically found to be stabilized in the inverse tetragonal XA type structure having a negative formation energy \cite{Singh_jmmm_14, Jin_cap_19}, which was later experimentally verified in our recent report \cite{Srivastava_aip_20}. Interestingly, the theoretical studies did predict that the Cr$_2$--based Heusler alloys possess a fully spin-polarized band structure, which is highly desirable in spintronics \cite{Felser_springer_16, Gao_apl_13}. On the other hand, Cr-based compound such as Cr$_3$Al exhibits a ferrimagnetic (FIM) nature with experimentally observed 84\% spin polarization \cite{Li_jap_09}. Moreover, Cr$_3$Al films were investigated using neutron diffraction to explore the magnetic moment of the atoms at different sites \cite{Boekelheide_prb_12}. In case of Cr$_2$CoAl, the appreciable amount of spin polarization 68\% was realized in the compensated ferrimagnetic (CF) state \cite{Singh_jmmm_14}. Theoretical studies predicted that the effect of compensation leads to a decrease in the magnetic moment in the Cr$_2$CoAl sample where the Cr-Cr neighboring atoms have an antiparallel coupling, and the individual magnetic moments for the nonsymmetric spin structure for the atoms at different sites were found to be Cr1 (1.36~$\mu_{\rm B}$), Cr2 (-1.49~$\mu_{\rm B}$), and Co (0.30~$\mu_{\rm B}$) in this inverse Heusler alloy \cite{Jin_cap_19, Meinert_jmmm_13, Singh_jmmm_14}. However, the full compensation is not experimentally realized with zero moment in the Heusler samples; for example,  Mn$_3$Ga has a magnetic moment of 0.65~$\mu$$_{\rm B}$/f.u, and MnCoVAl possesses 0.07~$\mu$$_{\rm B}$/f.u \cite{Kurt_prb_11, Meinert_japdap_11}. 
	
	Since the net moment of the samples is mainly governed by the magnetic atoms, in case of inverse Heusler alloys the magnetic moment follows the Slater-Pauling (SP) relation as $\rm M_{t} =(\rm Z_{t}-24)$~$\mu_{\rm B}$, where Z$_{\rm t}$ is the total number of valance electrons in the unit cell of the alloy. Here, for the complete magnetization compensation, the value of Z$_{\rm t}$ must be equal to 24 according to the SP relation \cite{Galanakis_prb_02}. Recently, FCF behavior was reported in Cr$_2$CoAl and Cr$_2$CoGa experimentally as well as by ab-initio calculations \cite{Srivastava_aip_20, Galanakis_apl_11, Meinert_jmmm_13, Luo_pb_15}. Our recent report on Cr$_2$Co$_{(1-x)}$Cr$_x$Al indicates that the XA structure is stable in the single-phase and we observed the signature of the FCF state in the $x=$ 0--0.4 samples \cite{Srivastava_aip_20}. At the same time, the magnetization curves show a finite hysteresis loop and do not saturate with magnetic field at a temperature of 50~K and 300~K. The magnetization behavior as a function of temperature and magnetic field has been classified as antiferromagnetic and/or compensated ferrimagnetic \cite{Srivastava_aip_20}. However, the crucial fact about the Heusler alloys is the presence of antisite disorder, which can decrease the spin polarization \cite{Mukadam_prb_16, Orgassa_prb_1999, Graf_pssc_11}. Using neutron diffraction (ND) \cite{Chatterji_sd_06}, L$\acute{\rm a}$zpita {\it et al.} determined the atomic distribution in the NiMnGa alloy in the paramagnetic (PM) region \cite{Lazpita_prl_17}. In the same way, an antisite disorder was found in the Mn$_2$VGa sample between V and Ga atoms \cite{Kumar_jpcm_12}. Recently, we have determined the magnetic structure of Co$_2$CrAl sample using powder ND across the phase transition and found the perfect agreement with magnetization behavior \cite{Nehla_prb_19}. Umetsu {\it et al.} also used powder ND to investigate the site occupancies in few Co$_2$ based full Heusler alloys both in the FM and PM states \cite{Umetsu_jalcom_10}. Powder ND has also been used to study the antisite disorder and magnetic structure in Co$_2$ based Heusler alloys \cite{Svyazhina_jetp_13, Raphael_prb_02}.  Interestingly, a structural transition, i.e., abrupt change in the lattice parameters was observed by temperature dependent ND, which found to be in agreement with the magnetic phase transition in TbCo$_2$ \cite{Halder_prb_18}. Also, ND is sensitive to the appearance of AFM ordering, and the observed T$_{\rm N}$ was found to be consistent with the magnetization data of CuMnSb \cite{Regnat_prm_18} as well as in inverse Heusler alloys \cite{AryalJALCOM20}. The magnetic phase transition in Cr$_2$CoAl is expected to be around 750~K \cite{Jamer_jmmm_15}; however, to the best of our knowledge there are no reports on magnetization and/or ND measurements at high temperatures. Therefore, it is of vital importance to unveil the magnetic structure, phase transition and structural disorder in the Cr$_2$CoAl inverse Heusler alloy.  
	
	In this paper, we present a detailed analysis of the ND patterns of the Cr$_2$CoAl sample to determine the crystal structure and the microscopic magnetic behavior over the large temperature range from 100~K to 900~K across the magnetic phase transition. The Rietveld refinement reveals the inverse tetragonal structure and no measurable antisite disorder in the sample. The analysis of the magnetic phase gives the value of net ordered moment around 0.04(4) $\mu$$_{\rm B}$/f.u. at 100~K, which found to decrease with temperature and reaches almost zero at around 730~K [defined as the magnetic ordering temperature T$_{\rm MO}$]. Also, the intensity plot of the (110) peak shows a similar decrease with temperature till 730$\pm$5~K and then become almost constant. Moreover, the lattice parameters increase with an increase in temperature, and the slope of the curve changes near the T$_{\rm MO}$. A similar anamoly is observed in the thermal factor and thermal expansion coefficient at around T$_{\rm MO}$. Interestingly, we find a nearly FCF structure where the magnetic moment of Cr2 shows antiparallel alignment with the Cr1 as well as the Co spins. The FIM transition obtained from the fitting of temperature dependence of the magnetic moment is found to be consistent with the intensity variation of the (110) peak with temperature.

\section{\noindent ~Experimental Details}
	
	Polycrystalline Cr$_2$Co$_{(1-x)}$Cr$_x$Al ($x=$ 0, 0.2) samples were prepared by arc melting (CENTORR, Vacuum Industries, USA). The basic characterization of these samples has been reported in ref.~\cite{Srivastava_aip_20}. Powder ND experiments are performed at the high-intensity diffractometer Wombat~\cite{Studer_pb_06} and the high-resolution diffractometer Echidna~\cite{Liss_pb_06,Avdeev_jpc_18} at the OPAL research reactor at ANSTO, Australia, using a cylindrical sample holder. A wavelength of $\lambda$ = 1.633~\AA~ and 1.622~\AA~were selected with a Ge(113) and a Ge(335) monochromator at the instruments Wombat (300-900~K) and Echidna (100~K), respectively. The neutron diffraction patterns were scanned at various temperatures on heating from room temperature to 900~K in the vacuum furnace. The step size was taken as 0.125$^{\rm o}$ in the 2$\theta$ range between 25$^{\rm o}$-135$^{\rm o}$ at the Wombat diffractometer for the $x=$ 0 sample. The measured diffraction pattern is analyzed with the Rietveld refinement method implemented with the FullProf package~\cite{Carvajal_07_Fullprof} considering the fundamental aspects of full-width and half maximum and other reliable parameters of the diffraction peaks~\cite{Rodriguez_pb_93}. The magnetic configuration is generated with neutron powder diffraction using the basis irreducible representation (BasIreps) function. 
	
\section{\noindent ~Results and Discussion}
	
\begin{figure}[ht]
		\includegraphics[width=3.6in, height=2.6in]{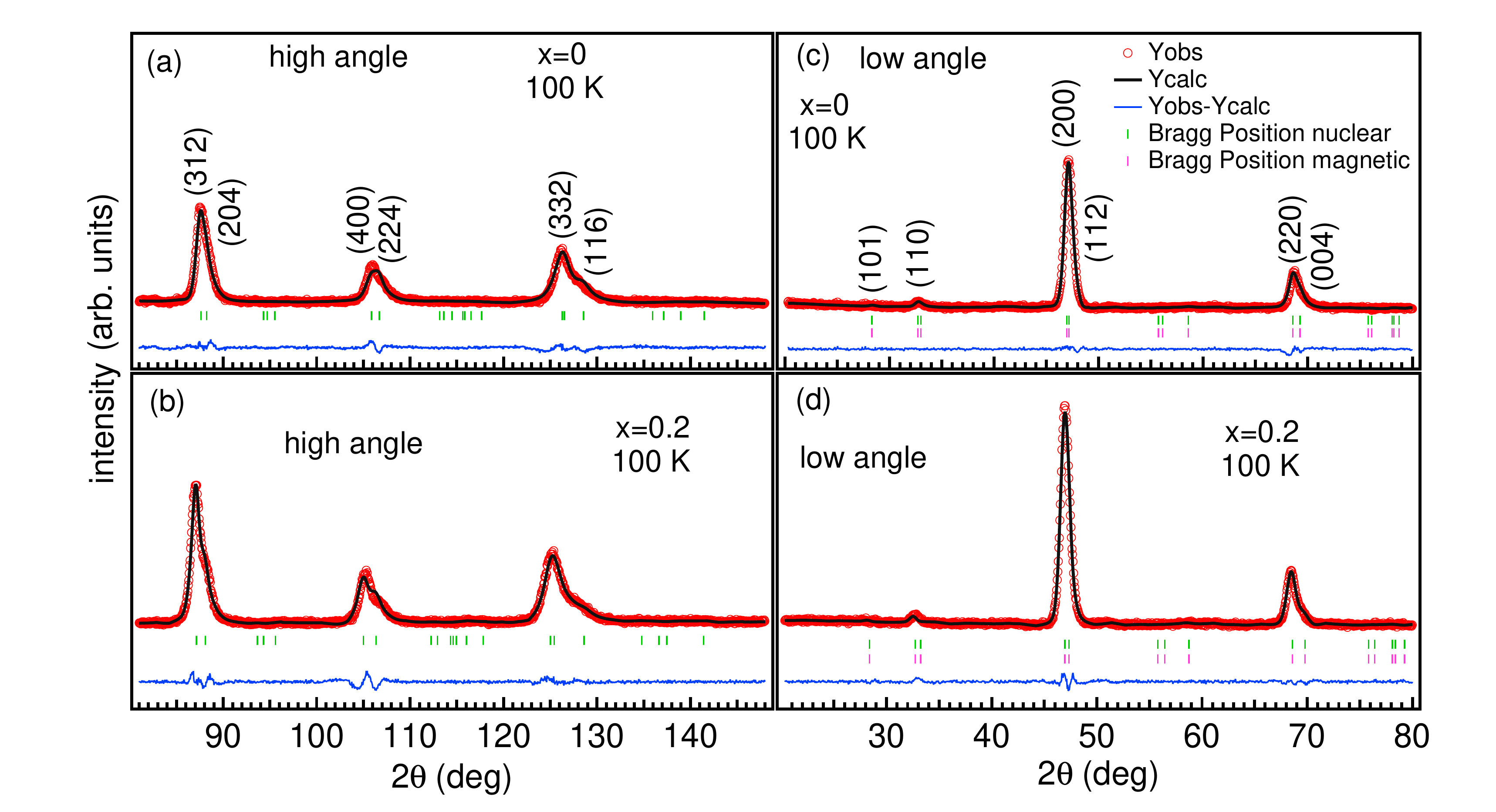}
		\caption{Rietveld refinement (black line) of the powder ND pattern (red symbols) (a, b) with a nuclear Bragg peaks at higher 2$\theta$ angles, and (c, d) with both nuclear and magnetic Bragg reflections at lower 2$\theta$ angles, measured at 100~K for both the $x=$ 0 and 0.2 samples. The difference profile (blue line) and Bragg peak positions (short vertical bars green for the nuclear and magenta for the magnetic) are shown in each panel. These high resolution patterns were recorded at the Echidna diffractometer ($\lambda$ = 1.622~\AA).}
		\label{fig-Fig_1}
	\end{figure}

	In Fig.~\ref{fig-Fig_1} we present the high resolution ND pattern of the Cr$_2$Co$_{(1-x)}$Cr$_x$Al ($x=$ 0, 0.2) samples measured on the Echidna diffractometer in the broad angle range 20$^{\rm o}$-150$^{\rm o}$ at 100~K. At first glance, we clearly observe the tetragonal distortion in the principal reflections for both the samples, 
which is consistent with the x-ray diffraction (XRD) patterns reported in ref.~\cite{Srivastava_aip_20}. The measurement temperature of 100~K was chosen to be in the magnetic region, as confirmed in the magnetization data \cite{Srivastava_aip_20}. In order to extract the information from the data at 100~K, we refine the ND pattern following the similar procedure as reported in ref.~\cite{Ahmed_jlcom_19}. First, the neutron powder-diffraction patterns at a higher angle (2${\rm \theta}$: 80--150$^{\rm o}$) have been refined as the magnetic form factor generally is negligible at higher angles above $\approx$80$^{\rm o}$ \cite{Kumar_jpcm_12, Ahmed_jlcom_19}. The refined pattern using the tetragonal structure with the space group I$\bar 4${\it m}2 considering the nuclear contribution only \cite{Sanjay_natcomm_16} are shown in Figs.~\ref{fig-Fig_1}(a, b) for the $x=$ 0 and 0.2 samples, respectively. We find the lattice parameters for the $x=$ 0 sample; $a=$ 4.051~\AA~ and $c=$ 5.665~\AA, and for the $x=$ 0.2 sample; $a=$ 4.075~\AA~ and $c=$ 5.680~\AA, which are consistent with the reports in Refs.~\cite{Jamer_jmmm_15, Srivastava_aip_20}. 

The ND data in the AFM state either show new Bragg peaks, which appear towards lower 2$\theta$ angle in the magnetically ordered state (below N\'eel temperature) or with the primitive lattice where the magnetic atoms arrange in such way that their multiplicity is higher than one, having a wave vector (k=0) \cite{McCalla_prm_21, Orlandi_prb_20, Kumar_jpcm_18}. However, in the present case, the magnetic atoms arrange in the multiplicity of two, which is higher than the multiplicity of the atoms in the primitive lattice. Since no additional Bragg peaks appear in the magnetically ordered state in our ND patterns, an AFM structure can be ruled out. This indicates either ferromagnetic (FM) or ferrimagnetic (FIM) ordering in these samples \cite{Mukadam_prb_16, Kumar_jpcm_18, Nehla_prb_19}. The magnetic contribution is associated with the (110) peak as the intensity of this peak increases at lower temperatures due to the presence of magnetic ordering \cite{Nehla_prb_19}. In order to reveal the magnetic structure, the low angle diffraction patterns are analyzed incorporating the magnetic contributions and using the lattice structure obtained from the high angle patterns, as shown in Figs.~\ref{fig-Fig_1}(c, d) for the $x=$ 0 and 0.2 samples, respectively. The extracted net ordered magnetic moments of the $x=$ 0 and $x=$ 0.2 samples are found to be 0.04(4) and 0.05(4)~$\mu$$_{\rm B}$/f.u. at 100~K, respectively. These values are reasonably in agreement with the generalized SP rule considering the total number of valence electrons in the unit cell \cite{SkaftourosPRB13, WurmehlJPCM06} as well as the value reported in Ref.~\cite{XiePRM22} using the band structure calculations. The antiparallel alignment and the different magnitude of the magnetic moment vectors of Cr and Co atoms indicate a nearly FCF structure (discussed later), which is consistent with the reported physical nature of FCF for the Cr$_2$CoAl sample in Ref.~\cite{XiePRM22}. All the extracted parameters for the $x=$ 0 sample are listed in Table~I of the Supplementary Information \cite{SI}. Notably, neutron diffraction was also used to study the FCF nature in the Mn$_2$V$_{1-x}$Co$_{x}$Ga Heusler alloys \cite{MidhunlalJPCM22}. 
		
		\begin{figure}
		\includegraphics[width=3.6in, height=5.3in]{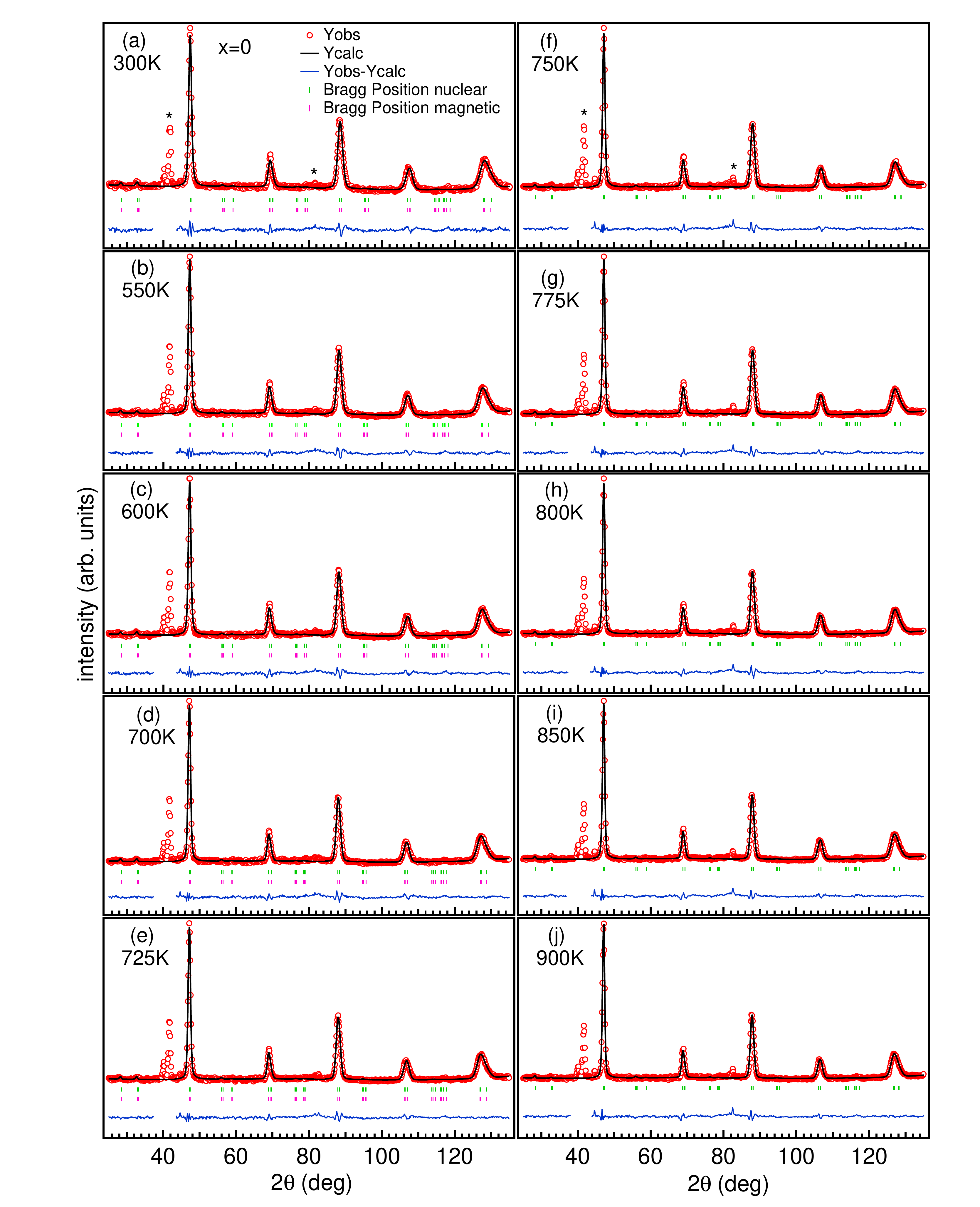}
		\caption{(a-j) The Rietveld refined neutron powder diffraction pattern of the $x=$ 0 sample, recorded on the diffractometer Wombat ($\lambda$ = 1.633~\AA). Each pattern is fitted with magnetic plus nuclear phases in the low temperature range 300--725~K, and with the nuclear phase only in 750--900~K range. The black asterisk tag indicates the peaks from the Niobium sample environment. The region 38$^{\rm o}$--43$^{\rm o}$ has been removed from the difference profile (blue line) for clarity in the presentation.}
		\label{fig-Fig_2}
	\end{figure}	
		
	Now, we mainly focus on the detailed analysis of powder ND pattern of the $x=$ 0 sample, collected at the diffractometer Wombat in the large temperature range from 300~K to 900~K, to reveal the magnetic structure and transition temperature. Figs.~2(a-j) show the Rietveld refined ND patterns considering magnetic plus nuclear phases in the temperature range of 300--725~K, and with only the nuclear phase from 750~K to 900~K range. There are a few peaks associated with the Niobium sample environment, between 2$\theta$= 38$^{\rm o}$--42$^{\rm o}$, as well as at $\approx$81$^{\rm o}$, which are present at all temperatures in Fig.~\ref{fig-Fig_2}. Therefore, for the sake of accuracy of the fitting parameters, these regions are excluded from the refinement by adjusting the range limit in the Fullprof program \cite{Carvajal_07_Fullprof}. Normally the ND technique is more sensitive as compared to XRD to quantify the antisite disorder due to the distinctly different neutron-bound scattering amplitude of the elements Cr (3.6~fm), Co (2.5~fm), and Al (3.5~fm) \cite{Nehla_prb_19, Neutron_92}. Therefore, we tried to find the antisite disorder between the Co and Cr atoms as well as between the Co and Al atoms by refining the ND patterns, initially at 900~K (above T$_{\rm{C}}$) with the nuclear phase only, as shown in Fig.~\ref{fig-Fig_2}(j). A similar method was reported to quantify the antisite disorder in the Mn$_2$VGa and Co$_2$MnSi Heusler alloys without affecting the stoichiometry where the atoms also have different scattering factors \cite{Kumar_jpcm_12, Raphael_prb_02}. However, we find no significant improvement in the refinement, which indicates the absence of measurable antisite disorders. On the other hand, the observed disorder between Cr and Al  by XRD analysis in Ref.~\cite{Srivastava_aip_20} cannot be ruled out from the ND analysis due to their similar neutron scattering cross-sections \cite{Neutron_92}. We also note here that any disorder between Cr and Al is not expected to affect the magnetic moment of these types of samples as predicted in Ref.~\cite{MiuraPRB04}. Therefore, the refined crystal structure inferred from the ND pattern measured at 900~K is used for the further analysis of the successive ND pattern at lower temperatures, as in Ref.~\cite{Lazpita_prl_17}.
		
\begin{figure}[ht]
		\includegraphics[width=3.6in, height=6.0in]{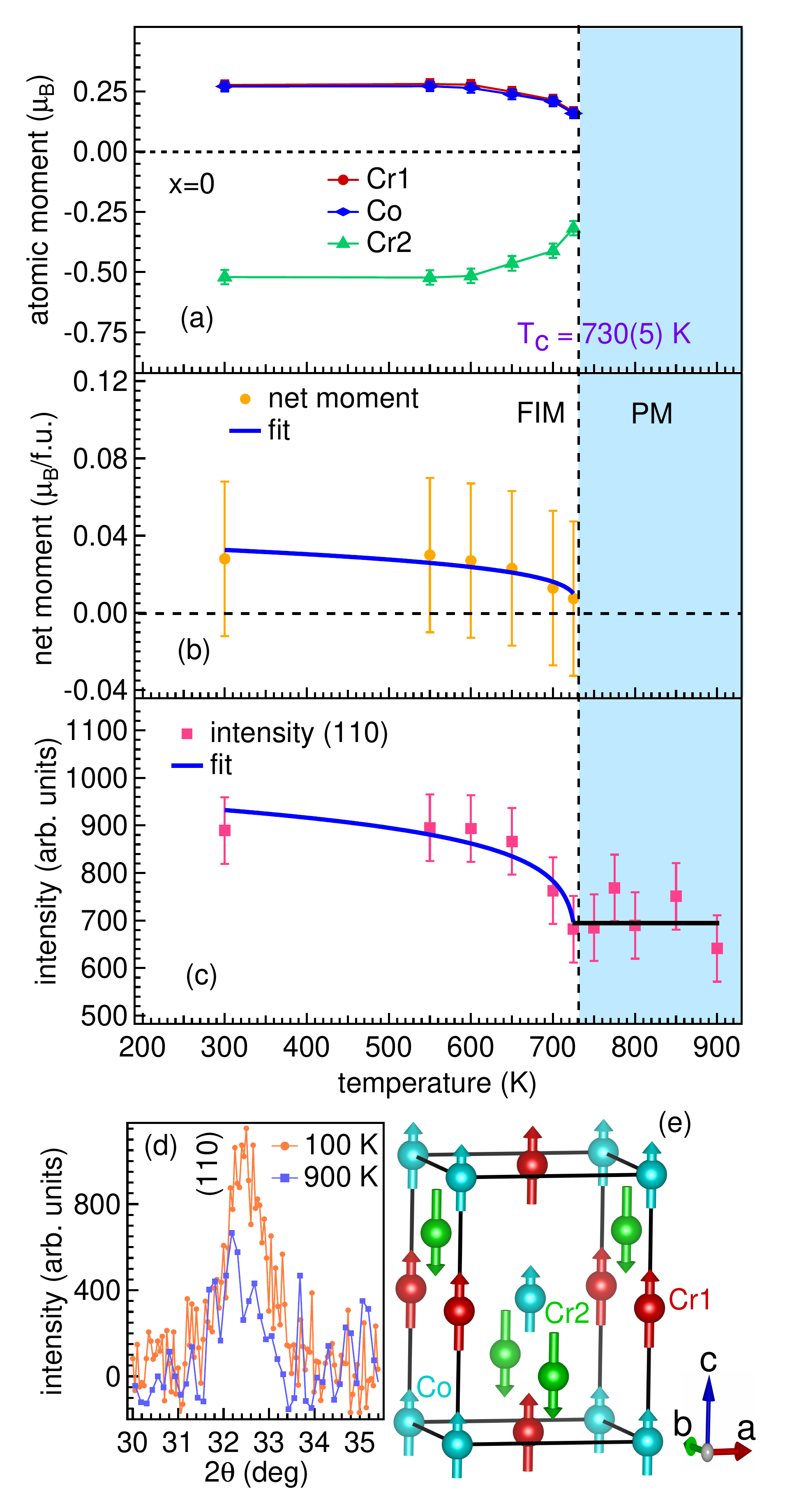}
		\caption{(a) The temperature dependence of the ordered magnetic moments of Cr1, Cr2 and Co sites. (b) The ordered net moment in the temperature range of 300~K to $\approx$730~K, and (c) temperature dependence of intensity of the (110) reflection for the $x=$ 0 sample. The blue solid lines are the fitted curves of magnetic moment and intensity with the power law equation. (d) The intensity of peak (110) at 100~K and 900~K for comparison, (e) the magnetic spin configuration with concordant ordering wave vector $k$ = (0, 0, 0) along the c-axis in the magnetic unit cell for 100~K. Here, the ND pattern at 100~K is recorded at the Echidna diffractometer ($\lambda$ = 1.622~\AA) and at high temperatures between 300~K and 900~K are recorded on the diffractometer Wombat ($\lambda$ = 1.633~\AA).}
		\label{fig-Fig_3}
	\end{figure}

	In order to analyze the neutron diffraction pattern in the magnetic region (below $\approx$750~K), it should be noted that there are no additional Bragg reflections in the magnetically ordered state of Cr$_2$CoAl Heusler alloy, see Fig.~2. However, with decreasing sample temperature the scattering intensity of the (110) peak increases, as plotted in Fig.~\ref{fig-Fig_3}(c), which suggests that the magnetic structure is either FM or FIM at low temperatures and excludes the possibility of a long-range AFM order in this $x=$ 0 sample \cite{McCalla_prm_21, Orlandi_prb_20, Kumar_jpcm_18, Kumar_jpcm_12}. Thus, to understand the magnetic structure and phase transition, we generate the magnetic moment configuration output using BasIREPS in the Fullprof program by considering the space group I$\bar{4}${\it m}2 and the magnetic state of FM or FIM. There are three magnetic atoms Cr1, Cr2, and Co and their corresponding Wyckoff positions are 2$b$(0, 0, 0.5), 2$d$(0, 0.5, 0.75), and 2$a$(0, 0, 0), respectively \cite{Jin_cap_19}. The appropriate magnetic propagation wave vector $k$ = (0, 0, 0) is considered for the FIM state with the best value by using the $k$-search option in the Fullprof program \cite{Carvajal_07_Fullprof}. This method provides the irreducible representation with only one basis vector $\Gamma$$_4$, which is related to the FIM interactions with real and imaginary positions as (0, 0, 1) and (0, 0, 0), respectively \cite{Moussa_prb_96}. The basis function helps to reveal the magnetic structure, where the arrangements of the magnetic moments are parallel or anti-parallel \cite{Ahmed_jlcom_19, Kumar_jpcm_18}. To extract the precise values of the magnetic moments from the ND pattern, it must be noted that the refinement is performed with the particular magnetic site of the atoms rather than the individual sites of the disorder positions \cite{Lazpita_prl_17, Mukadam_prb_16}. For example, the moment of the magnetic atoms with disorder gives the average moment of that atom at different Wyckoff positions. In refining the moment values at Cr1, Cr2, and Co sites, the sizable moment is found to be related to the (110) reflection only. Also, within the experimental error bar the magnetic moments in $ab-$plane were too small to be determined.
		
	Interestingly, the direction of the magnetic moment of Cr2 is found to be opposite to the ${c}$-axis whereas the moments of Cr1 and Co are parallel. It was theoretically reported that the Cr atoms show the opposite polarity owing to their mutual antiparallel configurations \cite{Singh_jmmm_14, Jin_cap_19}. For the refinement of the diffraction pattern below 750~K, we have initially taken all the structural parameters extracted from high temperature (900~K) data, and then refined the positions of the magnetic moment sites of the atoms to get the accurate microscopic magnetic moment values at a particular site. However, due to the strong direct interaction of $d-$states between the neighboring atoms of nonequivalent Cr atoms, the antiparallel spin configuration leads to an almost zero net magnetic moments \cite{Jin_cap_19, Meinert_jmmm_13}. The reliability parameters obtained from the refinement of the diffraction pattern at 100~K are $\chi$$^{2}$ = 2.9, Bragg R-factor = 1.4, RF-factor = 1.8, and magnetic R-factor = 4.1, which proves the good quality of the refinement \cite{Ravel_prb_02}. The obtained magnetic moment values and lattice parameters from the refinement are plotted in Figs.~3 and 4, and discussed in detail to understand the magnetic properties and phase transition in Cr$_2$CoAl sample. The ordered magnetic moments of the Cr1, Cr2, and Co sites obtained from the refinement of the powder ND patterns are plotted in Fig.~\ref{fig-Fig_3}(a), and the ordered net magnetic moment is shown in Fig.~\ref{fig-Fig_3}(b), which mimic the magnetization behavior and that the magnetic ordering disappear at high temperatures that suggests a transition from paramagnetic to the commensurate FIM magnetic structure at around T$_{\rm{C}}=$ 730$\pm$5~K. We also note here that the observed T$_{\rm{C}}=$ 730$\pm$5~K value of Cr$_2$CoAl using neutron diffraction is found to be consistent with the magnetization study on a similar system, i.e., Cr$_2$CoGa thin films, reported in Ref.~\cite{JamerAPL16}. However, the authors also observed a significant change in the T$_{\rm C}$ value depending on the annealing treatment to the thin films \cite{JamerAPL16}.  
	
	\begin{figure}[h]
		\includegraphics[width=3.65in, height=6.05in]{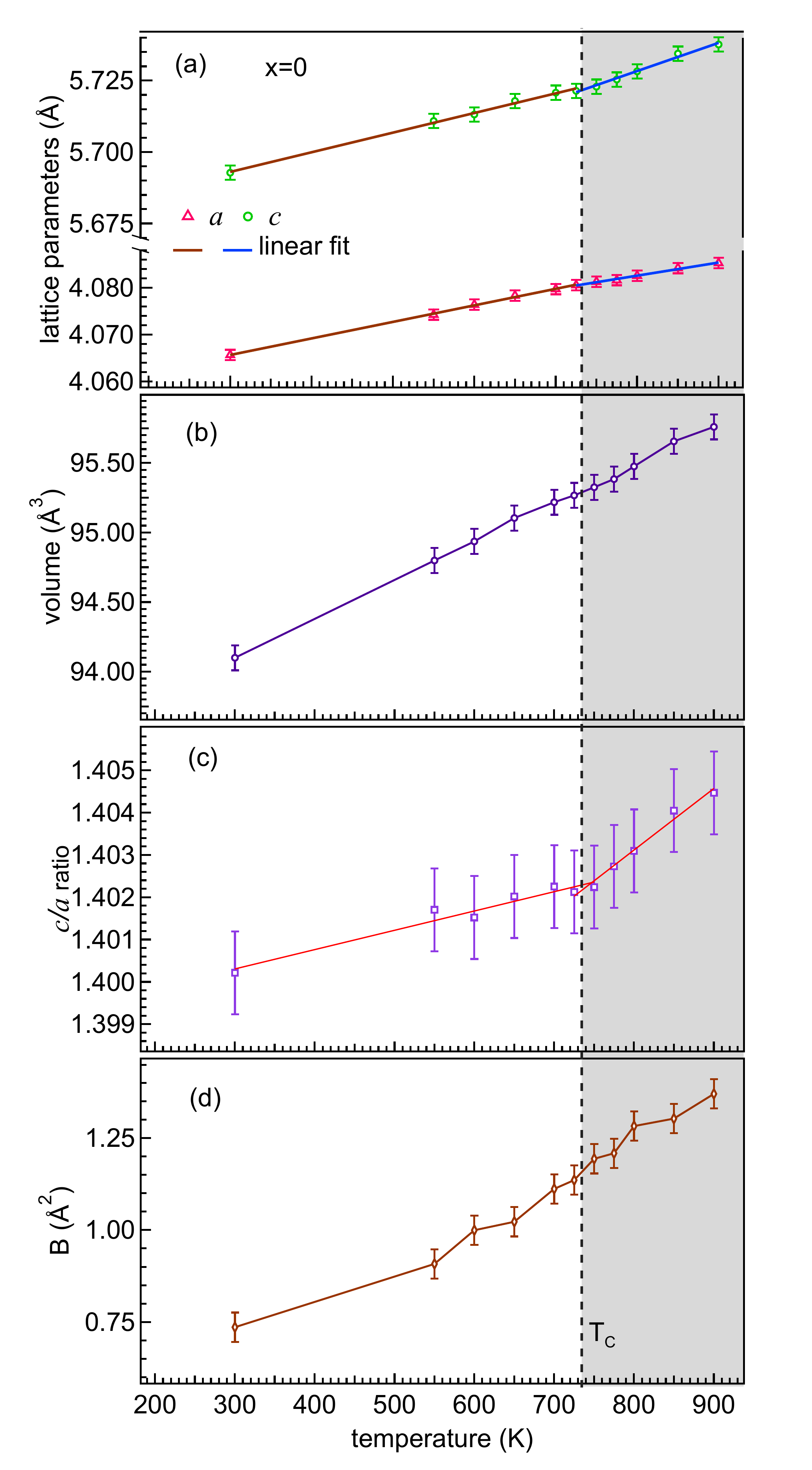}
		\caption{(a) The lattice parameters ($a$ and $c$) obtained from the Rietveld refinement and the solid brown ($\le$T$_{\rm C}$) and blue ($\ge$ T$_{\rm C}$) lines are the linear fit to the experimental data. (b) The variation in volume of the unit cell, and (c) the tetragonal ratio (${c/a}$) as a function of sample temperature. (d) The overall thermal factor (B) variation with the temperature obtained from the refinement. The black vertical dotted line shows the boundary of the ferrimagnetic transition. The error bars are standard deviations taken from the refinement.} 
		\label{fig-Fig_4}
	\end{figure}

In Fig.~\ref{fig-Fig_3}(c), the intensity of the (110) peak is plotted with temperature, which clearly increases below the transition temperature and is nearly constant in the PM region. The much higher intensity of the (110) peak observed at 100~K manifests the enhancement in the magnetic ordering at low temperatures. The ordered net magnetic moment and intensity of the (110) Bragg reflection versus temperature curves are analyzed by fitting an empirical power law: ${\rm M(T)=M_{0}\left(1-T / T_{\mathrm{C}}\right)^{\beta}}$ to the experimental data to determine the transition temperature \cite{Reehuis_prb_12, Taheri_prb_16, Orlandi_prb_20, Zhu_prb_15}. The FIM transition temperature (T$_{\rm C}$) is found to be 730$\pm$5~K with a critical exponent $\beta$= 0.2$\pm$0.1, which is well concordant with the critical exponent of the standard universality classes, assists to get the transition temperature where the intensity approach to zero with increasing temperature. The magnetic scattering is proportional to the square power of M \cite{Nehla_prb_19, Beleanu_prb_13}. In Fig.~\ref{fig-Fig_3}(d), the intensity of (110) peak at 100~K is observed $\approx$46\% higher than at 900~K, which manifests the magnetic ordering at low temperatures. Moreover, Fig.~\ref{fig-Fig_3}(e) shows the orientation of the moment vectors of the individual atoms in the magnetic unit cell at 100~K where the magnetic vectors of Cr2 are oppositely aligned with respect to the ${c}$-axis as well as to the moment vectors of Cr1 and Co atoms. This clearly reveals the nearly FCF order \cite{Ghanathe_jmmm_22} and is in good agreement with predictions from theoretical band structure calculations in Ref.~\cite{Jin_cap_19} as well as with other Cr based alloy \cite{VenkateswaraPRB18}. It is interesting to note that recently Xie {\it et al.} reported the FCF half-metallic nature in the inverse Heusler alloys that shows a spin polarized Weyl structure with quadratic nodal lines \cite{XiePRM22}.    
	
    Further, in Figs. \ref{fig-Fig_4}(a--d), we show the obtained lattice parameters ($a$ and $c$), unit cell volume, ${c/a}$ ratio, and overall thermal factor (B) inferred from the refinement of the ND patterns in the full temperature range for the $x=$ 0 sample. Fig.~\ref{fig-Fig_4}(a) shows a linear increase in the lattice parameters with temperature. There is a signature of change in slope at $\approx$730~K for both $a$ and $c$. These findings manifest the clear increase in the tetragonal distortion at this temperature as reflecting from the $c/a$ ratio shown in Fig.~\ref{fig-Fig_4}(c). The overall thermal factor (B), i.e., the Debye-Waller factor is plotted in Fig.~\ref{fig-Fig_4}(d), which also shows an increasing trend with temperature. The value of overall thermal factor well concurs with reported for Co$_2$CrAl and Co$_2$MnSi at room temperature \cite{Ahmed_jlcom_19, Nehla_prb_19}. We find that the thermal expansion in the lattice parameters has two regions of variation where an anomaly is observed at around 730~K. The lattice parameters follow the Bose-Einstein statistics for thermal expansion; therefore, the obtained lattice parameters ($a$ and $c$) are fitted with a general straight line equation in the two different regions below and above the phase transition temperature. The linear thermal expansion coefficient is calculated using the equation $\alpha=\frac{1}{a}\left(\frac{\partial a_{\mathrm{T}}}{\partial \rm T}\right)$ \cite{Yan_jped_08,Beleanu_prb_13,Neibecker_prb_17}, where $\alpha$ represent the linear thermal expansion coefficient and $(\frac{\partial a_{\mathrm{T}}}{\partial \rm T}$) are the values of the slope for the lattice parameters ($a$ and $c$). The obtained values of $\alpha$ (per K) for $a$ are 0.9$\times$10$^{-5}$ and 0.7$\times$10$^{-5}$, and for $c$ are 1.2$\times$10$^{-5}$ and 1.7$\times$10$^{-5}$ below and above $\approx$730~K, respectively. The value of $\alpha$ is found to be lower for $a$ side than for $c$ side, which indicates the significant expansion on the $c$ axis. These values show an anomaly in $\alpha$ around T$_{\rm{C}}=$ 730$\pm$5~K, which is due to the different slope of the lattice parameters as a result of the distortion in the inverse tetragonal crystal structure around T$_{\rm{C}}$. Here, the $\alpha$ values for Cr$_2$CoAl are well matched with the similar Heusler alloys as reported in refs.~\cite{Yan_jped_08, Nehla_prb_19, Neibecker_prb_17, Beleanu_prb_13}. In general, the value of $\alpha$ for the alloys and engineering metals is positive and in the order of 4$\times$10$^{–5}$/K \cite{Bonisch_natcomm_17}.

\section{\noindent ~Conclusions}
	
	In summary, we have investigated the magnetic structure and phase transition of the inverse Heusler alloys Cr$_2$CoAl using powder neutron diffraction measurements in the large temperature range of 100--900~K. The Rietveld refinement of the diffraction pattern manifests the single-phase inverse tetragonal structure of both these samples. We find no significant antisite disorder between Cr and Co atoms. More importantly, the ferrimagnetic (FIM) ordering is revealed by the refinement of the magnetic sites using the space group I$\bar{4}${\it m}2 and the magnetic wave vector, $k$ = (0, 0, 0) in the magnetically ordered state where the direction of the moment vectors of Cr2 is opposite to the ${c}$-axis as well as the moments of Cr1 and Co atoms. Interestingly, the net ordered magnetic moment as a function of temperature reveals the FIM ordering in the sample and the transition temperature is found to be 730$\pm$5~K. Moreover, we find the anomaly in the variation in the lattice parameters and the thermal expansion factor around the transition temperature, which can be attributed either to the magnetostriction or to the role of structural distortion in the magnetic phase transition in inverse Heusler alloys.  \\
	
\section{\noindent ~Acknowledgments}
	
	This work was financially supported by the BRNS through a DAE Young Scientist Research Award to RSD with Project Sanction No. 34/20/12/2015/BRNS. GDG thanks the MHRD, India, for fellowship through IIT Delhi. RSD gratefully acknowledges the financial support from the Department of Science \& Technology (DST), India, through the Indo-Australia Early and Mid-Career Researchers (EMCR) fellowship, administered by INSA (Sanction Order No. IA/INDO-AUST/F-19/2017/1887) for performing the neutron measurements at ANSTO, Australia. C.U. thanks the Australian Research Council for support through Discovery Grant No. DP160100545.

\end{document}